\documentclass{emulateapj}

\usepackage{hyperref}
\usepackage{url}
\usepackage{breakurl}
\usepackage{graphicx}
\usepackage{epstopdf}
\usepackage{longtable}
\usepackage{color}
\usepackage{float}

\newcommand{\myemail}{yksheen@kasi.re.kr}

\shorttitle{Recent Star Formation of Post-Merger Galaxies}
\shortauthors{Sheen et al.}

\begin{document}


\title{Recent galaxy mergers and residual star formation of red sequence galaxies in galaxy clusters}


\author{Yun-Kyeong Sheen\altaffilmark{1,2}, Sukyoung K. Yi\altaffilmark{3}, Chang H. Ree\altaffilmark{1}, Yara Jaff\'e\altaffilmark{4}, Ricardo Demarco\altaffilmark{2} and Ezequiel Treister\altaffilmark{5}}


\altaffiltext{1}{Korea Astronomy \& Space Science Institute, Daejeon, 305-348, Korea; \myemail}
\altaffiltext{2}{Departamento de Astronom\'ia, Universidad de Concepci\'on, Casilla 160-C, Concepci\'on, Chile}
\altaffiltext{3}{Department of Astronomy and Yonsei University Observatory, Yonsei University, Seoul 120-749, Korea; yi@yonsei.ac.kr}
\altaffiltext{4}{ESO, Alonso de Cordova 3107, Vitacura, 7630355 Santiago, Chile}
\altaffiltext{5}{Instituto de Astrof\'isica, Facultad de F\'isica, Pontificia Universidad Cat\'olica de Chile, Casilla 306, Santiago, Chile}


\begin{abstract}
This study explored the GALEX ultraviolet (UV) properties of optical red sequence galaxies 
in 4 rich Abell clusters at $z \leq 0.1$. In particular, we tried to find a hint
of merger-induced recent star formation (RSF) in red sequence galaxies. 
Using the NUV $-~r^{\prime}$ colors of the galaxies, RSF fractions were
derived based on various criteria for post-merger galaxies and normal galaxies. 
Following $k-$correction, about 36\% of the post-merger galaxies were classified as 
RSF galaxies with a conservative criterion (NUV $-~r^{\prime} \leq 5$),
and that number was doubled ($\sim$ 72\%) when using a generous criterion 
(NUV $-~r^{\prime} \leq 5.4$). The trend was the same when we restricted 
the sample to galaxies within $0.5\times$R$_{200}$. Post-merger galaxies
with strong UV emission showed more violent, asymmetric features in the deep optical images. 
The RSF fractions did not show any trend along the clustocentric distance within R$_{200}$. 
We performed a Dressler-Shectman test to check whether the RSF galaxies had any correlation 
with the sub-structures in the galaxy clusters. Within R$_{200}$ of each cluster, 
the RSF galaxies did not appear to be preferentially related to the clusters' sub-structures. 
Our results suggested that only 30\% of RSF red sequence galaxies show morphological 
hints of recent galaxy mergers. This implies that internal processes (e.g., stellar mass-loss
or hot gas cooling) for the supply of cold gas to early-type galaxies may play a significant role 
in the residual star formation of early-type galaxies at a recent epoch. 
\end{abstract}





\section{Introduction}

The origin of residual star formation in early-type galaxies is still under debate. 
Since the launch of the GALEX (Galaxy Evolution Explorer) ultraviolet (UV) space telescope 
\citep{mar05}, it has been discovered that about 30\% of massive early-type galaxies at 
$z <$ 0.1 show a hint of recent star formation (RSF) at a level of 1--3\% of their 
stellar mass \citep{kav07}. Galaxy mergers have been suggested as a primary 
driver of this phenomenon \citep{yi05,kav07}. Another reason
may be gas cooling in elliptical galaxies \citep{mat03,val15}. However, due 
to the difficulties in detecting signs of galaxy mergers and gas cooling in 
early-type galaxies, the residual star formation in early-type galaxies has only been 
investigated in detail for a few individual galaxies equipped with deep optical or X-ray
images \citep{fab95,osu01}. Several studies have therefore highlighted the need for 
deep optical imaging surveys of early-type galaxies \citep{kav09,sal10}.  

In \citet{she12}, post-merger galaxies were identified among massive red 
sequence galaxies (M$_{r^{\prime}} < -20$) in rich Abell clusters at 
$z\lesssim0.1$ using deep optical images.
The galaxies' features suggested that they had gone through galaxy mergers
relatively recently ($z \lesssim$ 0.5). Although their optical colors 
indicated that they were dominated by old stellar populations, as in
typical early-type galaxies, their post-merger features 
suggested that they may have had a certain level of star formation induced 
by recent merger events.

We explored the GALEX UV-optical colors of red sequence galaxies 
in Abell 119, Abell 2670, Abell 3330, and Abell 389 
at $z =$ 0.044, 0.076, 0.089, and 0.112, respectively, in order
to investigate the relation between recent galaxy mergers and 
the residual star formation of red sequence galaxies in galaxy clusters. 
One of the most effective ways to investigate recent 
star formation is to measure UV light 
from the galaxies, as the UV light is very sensitive to the existence of 
stellar populations younger than 1 Gyr. By combining deep UV and optical 
images of the four rich Abell clusters, a data set providing 
robust clues on mass-assembly histories as well as on the 
star formation histories of red sequence galaxies was established.

The UV-optical data and galaxy samples are described in Section 2.
In Section 3, the RSF fractions for post-merger galaxies
and normal galaxies are derived under various conditions, including $k-$correction, RSF criterion, 
and distance from the cluster center. The morphological properties, stellar populations 
and spatial distributions of the RSF galaxies are presented in Section 4,
and the results are discussed in Section 5.

\section{Data and Galaxy Samples}
 
GALEX UV images of three clusters (A2670, A3330, and A389) were taken in 
Deep Imaging Survey (DIS) mode to study UV upturn phenometnon in early-type galaxies 
\citep{ree07}. A119 imaging was performed in Medium Imaging Survey (MIS) mode. 
The GALEX exposure times for the cluster samples are presented in Table~\ref{uvexp}. 
UV catalogs of the target clusters were obtained from the GALEX GR7 Data Release. 
The Galactic extinction was corrected for GALEX FUV and NUV magnitudes using the
formulas $A_{\mathrm{FUV}} = 8.376 \times~E(B-V)$ 
and $A_{\mathrm{NUV}} = 8.741 \times E(B-V)$, as in \citet{wyd05}.

\begin{deluxetable}{llll}
\tablewidth{0pt}
\tablecaption{GALEX Data Properties \label{uvexp}}
\tablehead{
\colhead{Cluster} & \colhead{EXP$_{FUV}$} & \colhead{EXP$_{NUV}$} & \colhead{NUV detection rates} \\
\colhead{}& \colhead{(hours)} & \colhead{(hours)} & \colhead{among RS$_{spec}$ (\%)}
}
\startdata
A119 & 0.8 & 0.8 & 79.1 (53/67)\tablenotemark{a} \\
A2670 & 6.0 & 8.6  & 84.4 (81/96)\\
A3330 & 6.3 & 16.7  & 85.5 (53/62)\\
A389 & 6.0 & 8.7 & 94.4 (57/61)\\ \hline
Total & ... & ... & 85.3 (244/286)
\enddata
\tablenotetext{a}{Galaxy counts are presented in the parentheses.}
\end{deluxetable}

\begin{figure}
\includegraphics[scale=0.5]{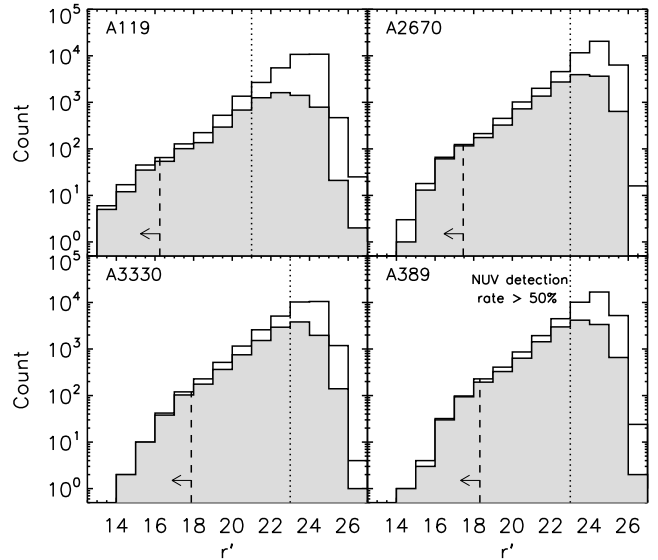}
\caption{Galaxy histograms from MOSAIC II $r^{\prime}-$band deep images 
for each cluster. The filled histograms show the counts of 
galaxies among the r-band galaxies detected in the GALEX NUV band.
The vertical dotted line distinguishes the magnitude bins with
NUV detection rates higher than 50\%. The magnitude 
limit for this study (M$_{r^{\prime}} < -20$) was calculated using the central 
velocity of each cluster, and is indicated with a dashed line. The study's galaxy samples 
are both above and below these limits, as marked with arrows.  \label{uvlimit}}
\end{figure}

Optical photometric catalogs were established from the deep optical images taken 
with the MOSAIC II CCD on the Blanco 4-m telescope at CTIO. The galaxy 
magnitudes in the $g^{\prime}, r^{\prime}$ bands were measured with the
Auto\_Magnitude (\texttt{MAG\_AUTO}) of the SExtractor \citep{ber96}, and their Galactic 
foreground extinction was corrected using reddening maps from \citet{sch98}.
The optical catalogs were then matched with the GALEX UV catalogs 
using a 6$\arcsec$ matching radius. Figure~\ref{uvlimit} presents 
galaxy histograms of MOSAIC II optical catalogs and the matched GALEX UV catalogs.
In the figure, the filled 
histograms represent galaxies matched with valid (note that GALEX
photometry failed to derive magnitudes on some detections) NUV magnitudes.
NUV detection 
rates were higher than 50\% for galaxies brighter than $r^{\prime} = 23$ in A2670, A3330, 
and A389, which were taken in DIS mode.  For A119, which had GALEX images in MIS mode,
the limit for a 50\% NUV detection rate was $r^{\prime} = 21$.  
To provide a guide for our galaxy samples, the magnitude limit (M$_r^{\prime} < -20$)
was approximately calculated using the distance modulus of each 
cluster (36.25, 37.46, 37.88, and 38.32 for A119, A2670, A3330, and A389, respectively)
and was indicated with a dashed line on the figure. 
Our galaxy samples had magnitudes above and below these limits, as
the absolute magnitude of each galaxy was calculated based on its own 
spectroscopic redshift. 

\begin{figure*}
\includegraphics[scale=0.5]{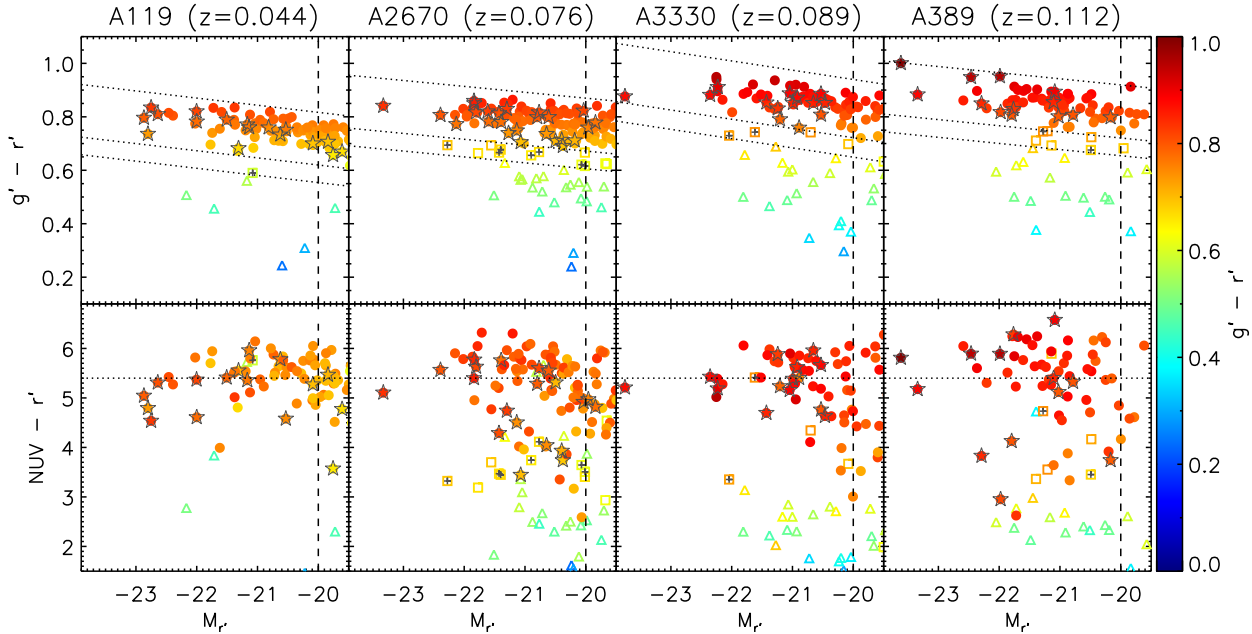}
\caption{UV -- optical color-magnitude relations (CMRs) of target clusters. 
Only the spectroscopic members are plotted. The symbols are color-coded 
according to their $g^{\prime} -~r^{\prime}$ colors. The vertical dashed lines mark 
the magnitude cut (M$_{r^{\prime}} = -20$) for the volume-limited samples. 
{\textit{Top:}} The optical red sequence, green valley, and blue clouds were defined 
with their color-magnitude relations of $g^{\prime} -~r^{\prime}$ vs
M$_{r^{\prime}}$, as represented by the dotted lines. We used different symbols 
for the red sequence (filled circles), green valley (open squares), and 
blue clouds (open triangles) in the diagrams. The post-merger galaxies identified 
by \citet{she12} among the red-sequence galaxies were indicated with superimposed open 
gray stars. The crosses mark green valley galaxies with 
post-merger signatures. {\textit{Bottom:}} UV -- optical CMRs of spectroscopic 
cluster members using GALEX UV and Blanco MOSAIC 2 optical images. 
The meanings of the symbols were the same as for the optical CMRs. The 
NUV $-~r^{\prime} = 5.4$ color cut was drawn as a RSF criterion adopted 
from the literature (e.g. \citet{yi05} among others). \label{uvcmr2}}
\end{figure*}

Optical red sequence galaxies were identified from the $g^{\prime} -~r^{\prime}$ vs 
M$_{r^{\prime}}$ color-magnitude relations (CMRs) of the spectroscopic members of
the clusters. As introduced in \citet{she12}, the cluster memberships were assigned using 
the velocity distributions from the spectra taken by a multi-object spectrograph, 
Hydra, on the Blanco 4-m telescope at CTIO. For A119 and A2670, the spectroscopic 
catalogs were supplemented with spectroscopic redshifts from SDSS (Sloan Digital Sky
Survey) for missing objects from Hydra observations.
The completeness of the spectroscopic survey for the massive red sequence galaxies 
(M$_{r^{\prime}} < -20$ using the distance moduli of the clusters)
were 94\%,  92\%, 81\%, and 75\% for A119, A2670, A3330, and A389, respectively.
The line-of-sight velocity dispersions ($\sigma_{los}$) were 895, 1039, 869, 873 km/s, respectively. 
Cluster memberships were assigned to galaxies within $\pm3\sigma_{los}$ of the
velocity distribution of each cluster.
In this paper, we only considered the cluster members with spectroscopic redshifts. 
Optical CMRs of the spectroscopic cluster members are presented in the top panels in 
Figure~\ref{uvcmr2}. We should mention that our Hydra observations were focused
to get cluster memberships of red sequence galaxies first. Therefore the Hydra redshift catalogs 
are not complete for galaxies in blue clouds.
To begin with, the red sequences were determined through 
iterative fitting using the 2$-\sigma$ clipping method. The red sequence, green 
valley, and blue clouds were then defined according to their $g^{\prime} -~r^{\prime}$ 
color ranges as $\pm~3\sigma$, $-~3\sigma \sim -~5\sigma$, and  
$< -~5\sigma$, respectively, for the best fit of each cluster
(the dotted lines in the top panels in Figure~\ref{uvcmr2}). 
We identified 67, 96, 62, and 61 red sequence galaxies from 
A119, A2670, A3330, and A389, respectively (including 4 galaxies and 22 galaxies 
from the SDSS spectroscopic redshift catalogs for A119 and A2670).
Therefore, our volume-limited red sequence sample included 286
galaxies with a magnitude limit of M$_{r^{\prime}} < -20$ from 
the four target clusters. Among the 286 red sequence galaxies, 
263 have valid NUV magnitudes from 
the GALEX catalogs.

Prior to UV analysis, we conducted a visual inspection of the photometric 
apertures used by the GALEX pipeline, as introduced in \citet{yi11}.
This process is critical to the avoidance of contamination from nearby UV bright objects 
that were not resolved into separate objects by the pipeline due to the low 
spatial resolution (1.5 arcsec pixel$^{-1}$) of GALEX images. 
Nineteen galaxies, including 3 post-merger galaxies,
were rejected from the 263 NUV-detected galaxies. 
The rejection rate was 7.2\% (19/263), which was similar to that found in \citet{yi11} for 
the early-type galaxies in clusters with SDSS data (6.8\% (88/1294)). 
Thus, we obtained 244 red sequence galaxies with robust NUV detections. 
Henceforth, these volume-limited red sequence galaxies with NUV magnitudes 
will be called ``RSVL'' (Red Sequence Volume-Limited) samples. The UV-optical color-magnitude relations 
of the target clusters are presented in the bottom panels of Figure~\ref{uvcmr2}.
The figure shows that the galaxies lying in a tight red sequence in the optical CMRs are widely spread
in UV-optical CMRs.

\section{RSF Fractions}
\label{sec:rsf}

\begin{figure*}
\center{
\includegraphics[scale=0.5]{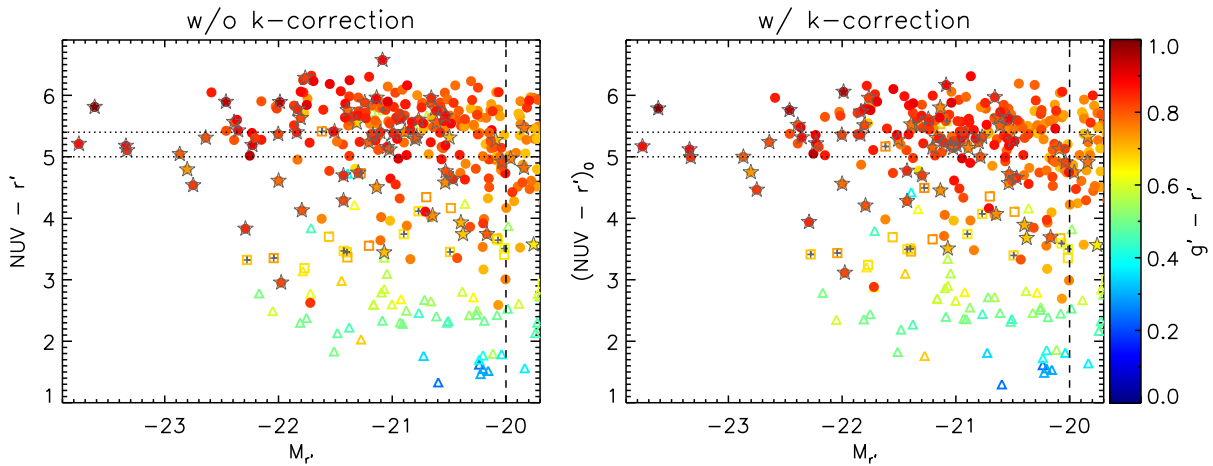}
\caption{
Combined UV-optical CMRs before and after $k-$correction of 
NUV $-~r^{\prime}$ colors (left and right). The symbols are the same as in 
Figure~\ref{uvcmr2}. The volume-limited sample cut of 
(M$_{r^{\prime}} = -20$) is represented by a dashed line. 
The conventional RSF cut (NUV $-~r^{\prime}$ = 5.4, 
the upper dotted line) was found to be rather generous. NUV $-~r^{\prime} = 5$ 
(the bottom dotted line) appeared to match the optical red-sequence galaxies better, 
especially for the $k-$corrected NUV $-~r^{\prime}$ colors.
\label{uvcmr2_all}}
}
\end{figure*}

In this work, we utilised NUV $-~r^{\prime}$ colors as the only criteria 
for identifying RSF galaxies among the red sequence galaxies.
However, as shown on
the optical CMRs in Figure~\ref{uvcmr2}, the optical color shifts of the red 
sequence galaxies between clusters at different redshifts demanded
a $k-$correction of the magnitudes. We computed the $k-$correction of 
the UV-optical magnitudes using two-component stellar population 
modeling. This involved matching the observed photometric SEDs 
(Spectral Energy Distributions) with composite model SEDs made 
up of young and old stellar populations. In our modeling, the old 
population was fixed as a 12 Gyr old stellar population. We 
constructed model SEDs using the models of {\it Starburst99} \citep{lei99} 
for young stellar populations and the models of \citet{yi03} for an old stellar population,
by varying the age and mass fraction 
of the young stellar component as well as the internal extinction. 
Both stellar population had a solar metallicity.
The best match between an observed SED and a model SED was found 
with a $\chi^{2}$ test. We derived the $k-$correction terms by comparing the 
magnitudes from the best fit model at the spectroscopic redshift of a galaxy 
and another one shifted to the rest frame. For more details on the 
two-component stellar population modeling, please refer to \citet{she09}. 
Figure~\ref{uvcmr2_all} shows the combined UV-optical CMRs of the 
four clusters before and after the $k-$correction.
In order to see the effect of the $k-$correction in this study with galaxies 
at $z \lesssim 0.1$ and to compare it with other studies that did not apply 
$k-$correction to the galaxy UV-optical colors, we derived the RSF fractions 
of both sub-samples with and without $k-$correction. In the following section, however, 
we discuss only the results of the $k-$corrected samples, unless stated otherwise.

One of the conventional selection criteria for quiescent galaxies has been NUV $-~r^{\prime} >$ 5.4, 
based on the NUV $-~r^{\prime}$ color of a representative UV upturn galaxy, 
NGC 4552 \citep{yi05,kav07}. \citet{jeo09} later suggested 
NUV $-~r^{\prime} \leq 5$ as a conservative limit for selecting RSF galaxies. 
\citet{cro14} tested the two criteria using post-starburst galaxies and found 
that a post-starburst galaxy appeared redder than NUV $-~r^{\prime} = 5$. 
According to Figure~\ref{uvcmr2_all}, however, the majority 
of red sequence galaxies appeared to reside at $5 <$ NUV $-~r^{\prime} <  6$ after 
$k-$correction. Since those NUV $-~r^{\prime}$ cuts were empirical suggestions, 
we present the results for both criteria in order to avoid a bias from ambiguous RSF criteria.

\def\arraystretch{1.5}
\begin{deluxetable*}{crrrr}
\tabletypesize{\footnotesize}
\tablewidth{0pt}
\tablecaption{The Average RSF Fractions (in percent) \label{avgfrac}}
\tablehead{
\colhead{Galaxy samples} & \multicolumn{2}{c}{w/ k-correction} & \multicolumn{2}{c}{w/o k-correction} \\ 
\hline
\colhead{} & \colhead{NUV$ -~r^{\prime} \leq$ 5.4} & \colhead{NUV$ -~r^{\prime} \leq$ 5} & \colhead{NUV$ -~r^{\prime} \leq$ 5.4} & \colhead{NUV$ -~r^{\prime} \leq$ 5}
}
\startdata
Post-mergers & 72.4$\pm$14.7& 36.2$\pm$9.2 &62.1$\pm$13.2 & 32.8$\pm$8.7  \\
Normal galaxies &50.0$\pm$6.4 & 26.3$\pm$4.2 & 41.9$\pm$5.7 & 22.6$\pm$3.9  \\
\hline
Post-mergers $<$ 0.5 R$_{200}$ & 70.0$\pm$17.2 & 30.0$\pm$9.9 & 60.0$\pm$15.5\% & 27.5$\pm$9.4  \\
Normal galaxies $<$ 0.5 R$_{200}$ & 51.1$\pm$7.5 & 25.9$\pm$4.8 & 43.2$\pm$6.7\% & 21.6$\pm$4.3  \\
\hline
All RSVL&55.3$\pm$5.9 &28.7$\pm$3.9  & 46.7$\pm$5.3&25.0$\pm$3.6  \\
All RSVL ($<$ 0.5 R$_{200}$) &55.3$\pm$6.9 &26.8$\pm$4.4 &46.9$\pm$6.2 &22.9$\pm$4.0 
\enddata
\end{deluxetable*}

\begin{figure}
\includegraphics[scale=0.47]{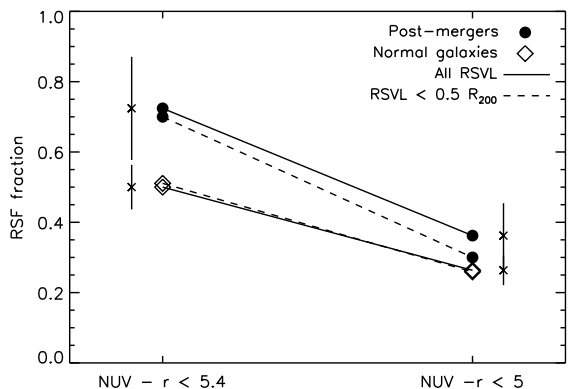}
\caption{We compared the average RSF fractions using $k-$correction (presented in Table~\ref{fraction}) 
with different NUV $-~r^{\prime}$ cuts, and either the full field-of-view sample or the central 
($< 0.5 R_{200}$) sample. The error bars are for the fractions with all RSVL samples 
(solid lines). \label{uvstat_all3_2}}
\end{figure}

The RSF fractions of the RSVL samples were calculated using different criteria based on 
the morphological signatures (post-mergers vs normal/featureless galaxies), 
$k-$correction ($k-$corrected vs not $k-$corrected), NUV $-~r^{\prime}$ 
cuts (NUV $-~r^{\prime} = 5.4$ or 5), and distances from the cluster center 
(RSVL samples in the full MOSAIC II field-of-view vs RSVL within 
$0.5\times$R$_{200}$ of each cluster).  Table~\ref{avgfrac} shows the
averages of the RSF fractions from the four Abell clusters (the fractions of 
each cluster are presented in Appendix~\ref{append}). 
On average, about 36\% of the post-merger galaxies showed RSF signatures 
with NUV $-~r^{\prime} \leq 5$. This was about 40\% larger than 
the fraction of the normal galaxies ($\sim$ 26\%).  
When applying NUV $-~r^{\prime} = 5.4$ as the RSF criterion, the fractions 
almost doubled in both sub-samples. The ratio of the fractions 
between the post-merger galaxies and the normal galaxies was $\sim$ 1.4 
regardless of the RSF criterion. Following $k-$correction, the galaxy colors typically became 
bluer in the UV-optical bands. Therefore, the RSF fraction was slightly higher after the
$k-$correction. Our target clusters covered slightly different areas in the optical data
as they were located at different redshifts within $z \lesssim$ 0.1. 
Therefore, we also checked the RSF fractions for galaxies within 0.5$\times$R$_{200}$ 
of each cluster. As shown in Table~\ref{avgfrac}, we could not find significant
differences in the average RSF fractions of the galaxies within 0.5$\times$R$_{200}$ 
as compared to the fractions using all RSVL samples. 

A schematic view of the RSF fractions is presented in Figure~\ref{uvstat_all3_2}. 
The RSF fractions are plotted along with the different RSF criteria for post-merger 
galaxies and normal galaxies. The solid lines are used to represent all RSVL samples and the 
dashed lines are for the samples within 0.5$\times$R$_{200}$. 
The post-merger galaxies showed a slight lack of UV bright (NUV $-~r^{\prime} \leq 5$)
galaxies in the central region. 
Although the errors are large, this makes sense, 
as most post-merger galaxies may have gone 
through galaxy mergers at the cluster outskirts.
The merger-induced star formation is probably quenched 
(by environment) and their stellar populations have aged during the dynamical friction 
time of the galaxy with respect to the cluster.

\section{Properties of RSF Galaxies}
\subsection{Morphology}
\label{sec:morph}

\begin{figure*}
\center{
\includegraphics[scale=0.8]{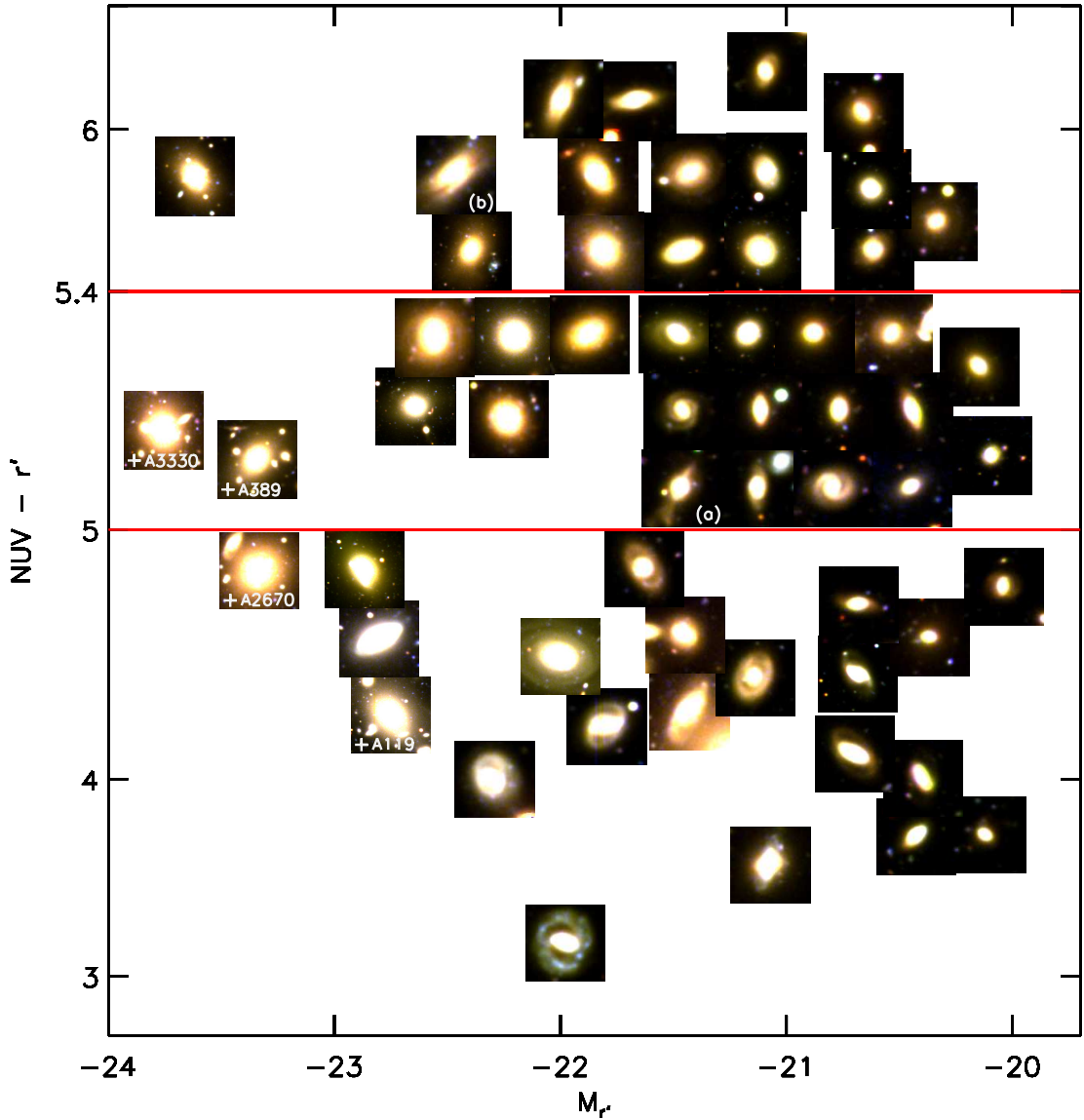}
\caption{Thumbnail images of post-merger galaxies in red sequences of 
A119, A2670, A3330, and A389. Red lines indicate the RSF criteria 
used in this paper (NUV $-~r^{\prime} = 5$ and $5.4$). 
The four BCGs are marked with crosses in the bottom-left corner
of the thumbnails. (a) is a galaxy with a violent post-merger feature in 
the color range of 5 $<$ NUV $-~r^{\prime} \leq$ 5.4. (b) shows a
dusty disk structure along with a faint structure which spreads beyond
the area of this thumbnail image. Please refer to Section 4.1 for those galaxies. 
The galaxys' positions in the diagram had to be adjusted 
to avoid overlaps, and the image scale for the brightest galaxies is different. 
Post-merger features were 
identified at \citet{she12} using disturbed features, e.g., asymmetric 
structures, faint features, discontinuous halo structures, rings and dust lanes.
The images have been provided to show the galaxies' overall 
optical colors and morphology rather than their post-merger features. 
The features are sometimes very faint and spread over a wide area, making it
complicated to show them in the color composite images. \label{cmd_uvthumb}}
}
\end{figure*}

\begin{figure*}
\center{
\includegraphics[scale=0.45]{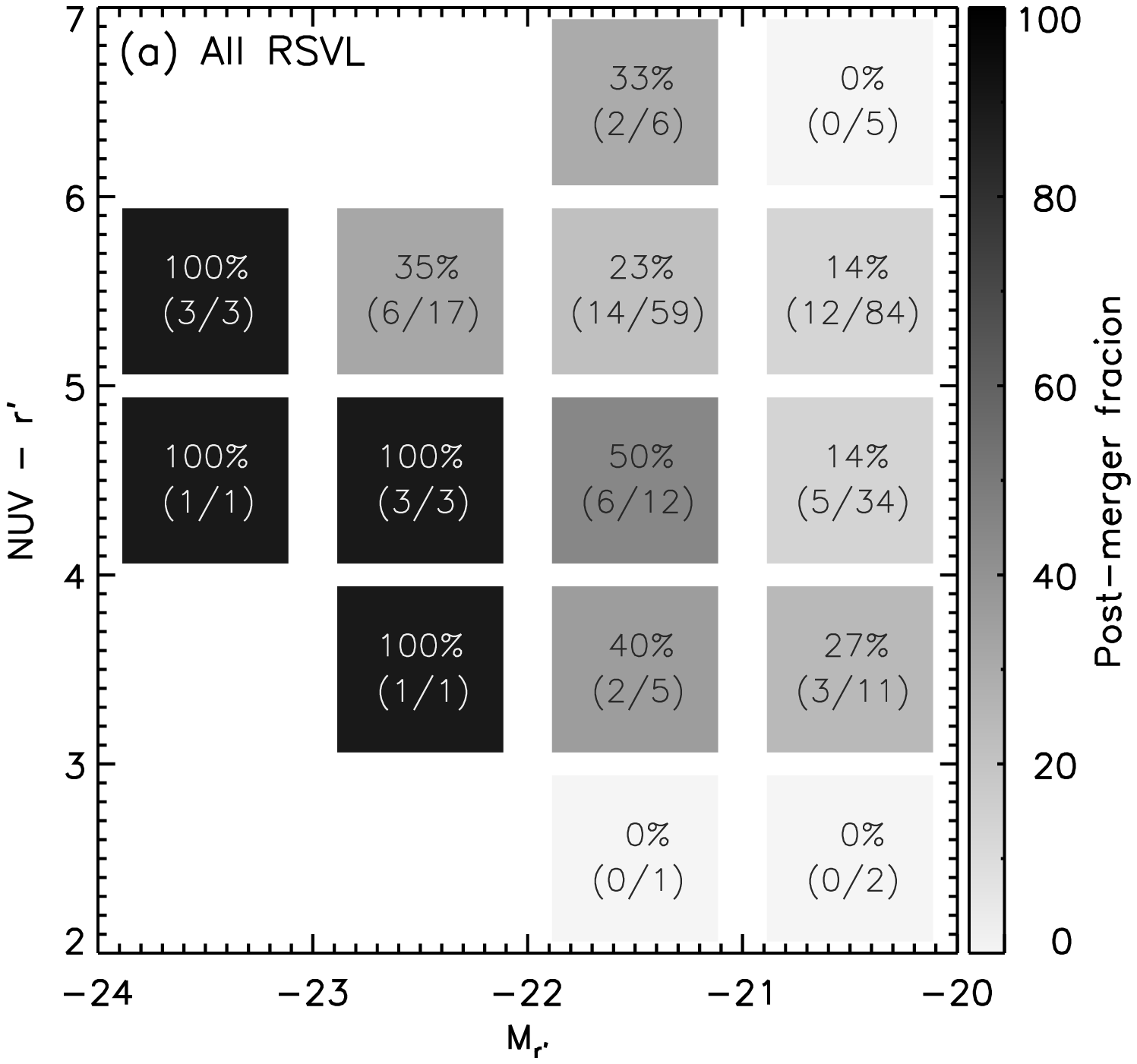}
\includegraphics[scale=0.45]{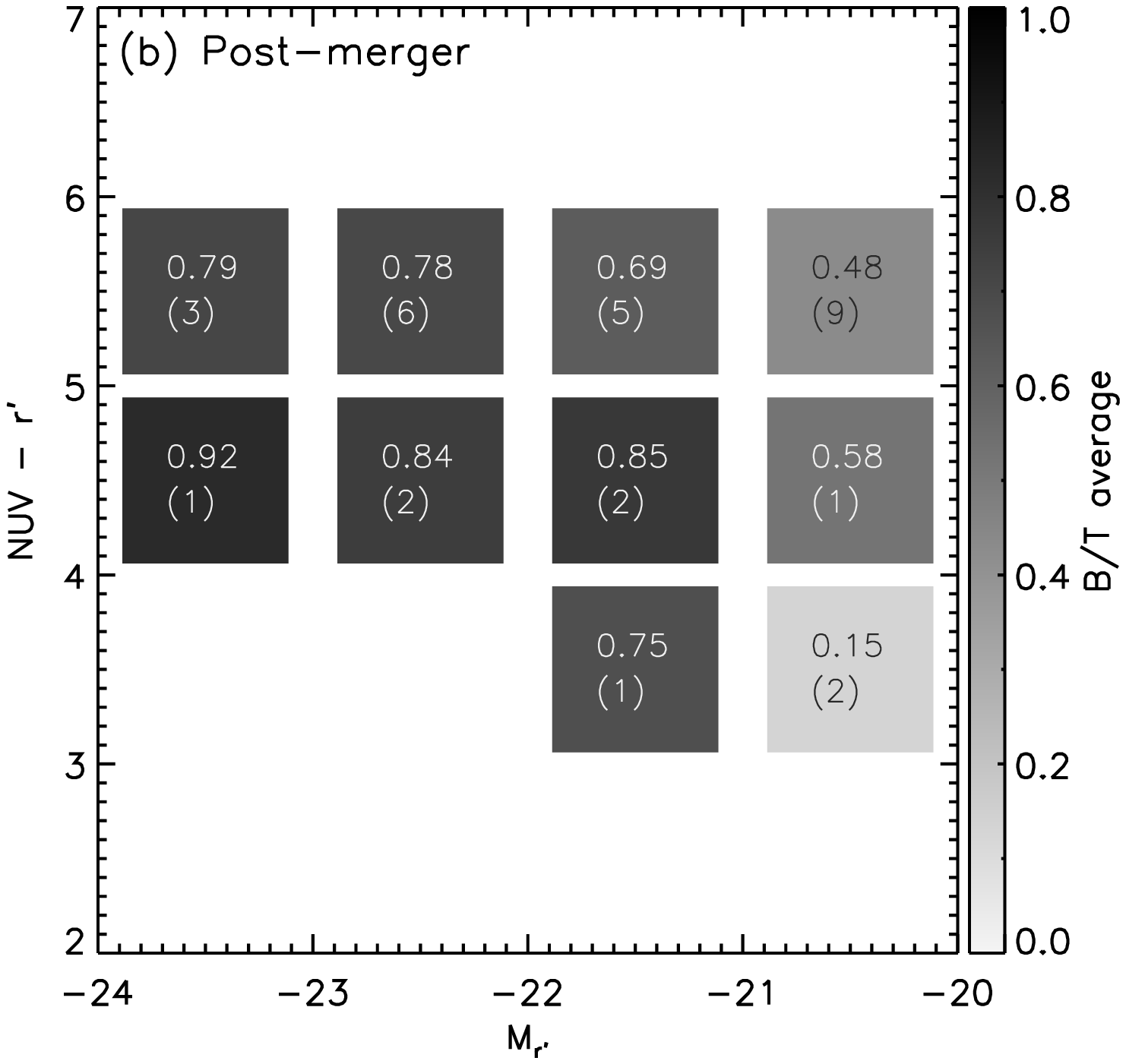}
\includegraphics[scale=0.45]{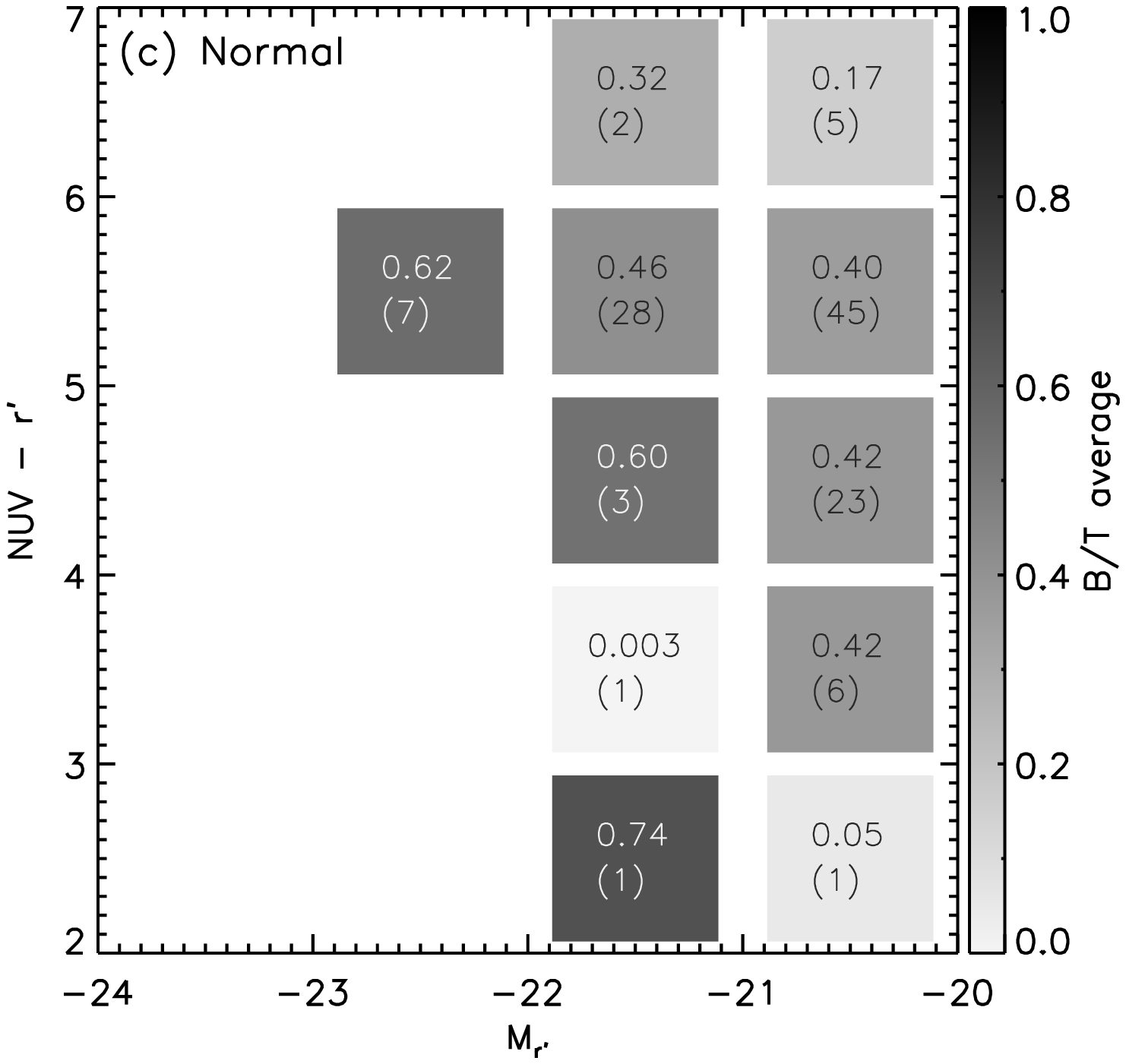}
\includegraphics[scale=0.45]{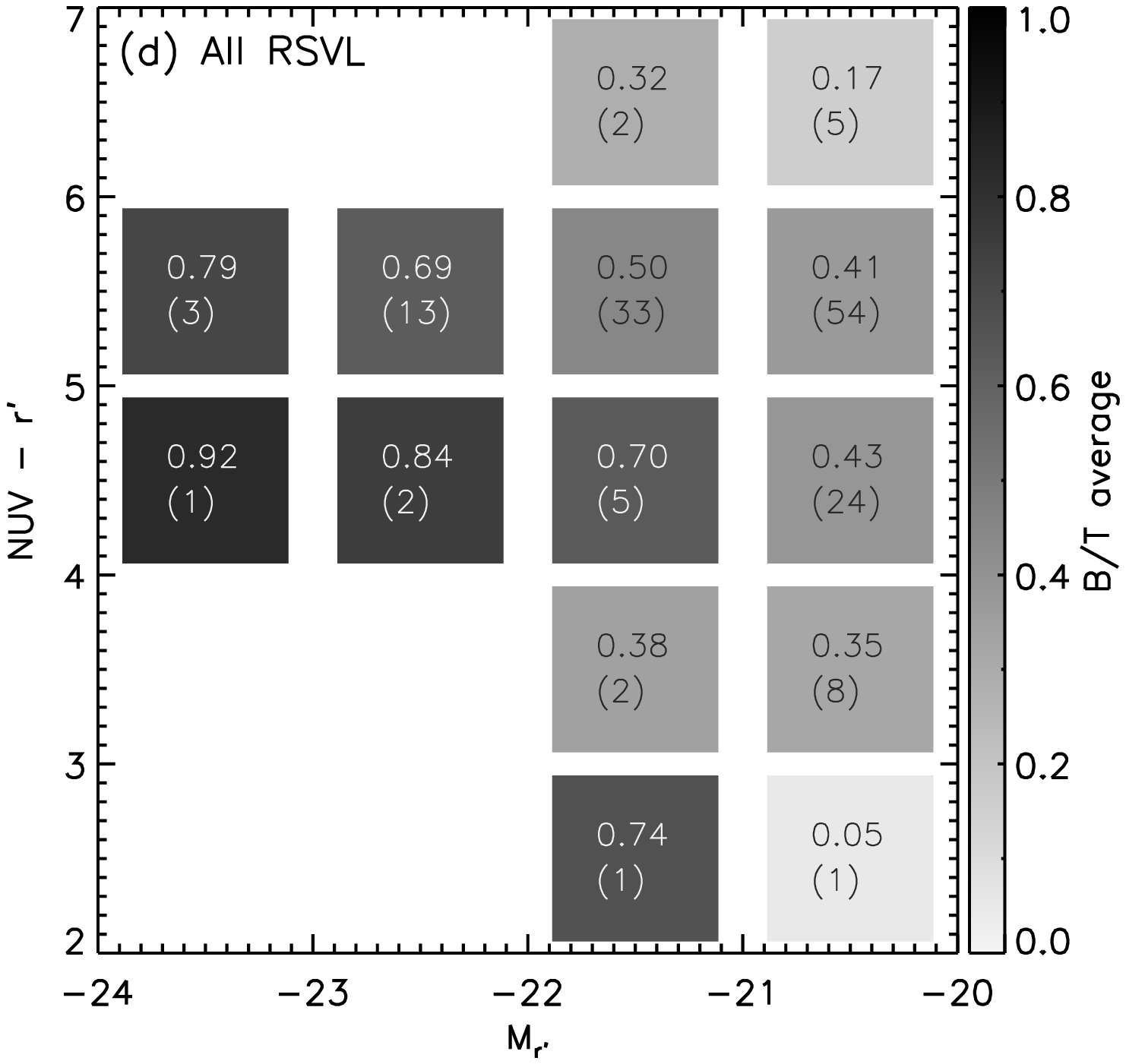}
\caption{(a) Post-merger fractions for red sequence galaxies
presented in grid of M$_{r^{\prime}}$ and  NUV $-~r^{\prime}$ colors. 
The numbers of post-merger galaxies and red sequence galaxies located 
in each bin are also given in parentheses. (b) Average bulge-to-total 
(B/T) ratios of post-merger galaxies calculated for each bin. Only 
galaxies with $\widetilde{\chi}^{2} \le 2$ were included in the calculation, 
and the number of galaxies is shown in the parentheses. (c) Average B/T 
ratios of normal, featureless galaxies. (d) Average B/T ratios for all red 
sequence galaxies ($\widetilde{\chi}^{2} \le 2$). \label{uvfrac}  }
}
\end{figure*}

The galaxy samples were selected based on their optical colors. 
The previous section, however, showed that UV-optical colors spanned 
a wide range. We considered the way the post-merger features differed 
according to the UV-optical colors. Post-merger features were 
identified at \citet{she12} using disturbed features, e.g., asymmetric 
structures, faint features, discontinuous halo structures, rings and dust lanes.

To begin with, we examined the appearance of post-merger galaxies.
Thumbnail images of the post-merger galaxies are presented 
within a plot of NUV $-~r^{\prime}$ colors vs M$_{r^{\prime}}$  in
Figure~\ref{cmd_uvthumb}. The galaxies were divided into three groups of
NUV $-~r^{\prime}$ colors based on the two RSF criteria used in 
this paper (red lines in the figure). 
We found that galaxies at NUV $-~r^{\prime} < 4$ displayed very 
asymmetric shapes and blue spots from recent star formation.
In the $4 <$ NUV $-~r^{\prime} < 5$ range, the galaxy 
morphologies appeared more symmetric, some showing thick bar structures 
or ring structures. Massive elliptical galaxies within this color range showed faint 
structures such as shells or filamentary structures in their halos. 
In the 5 $<$ NUV $-~r^{\prime} < 5.4$ range, blue spots were no longer
found in the galaxies. Most of them showed elliptical shapes, 
while some of them presented short spiral structures. A violently-disturbed 
galaxy that was likely to be a remnant of a 
galaxy merger between gas-poor galaxies was also found in this range ((a) in 
Figure 6).
In the NUV $-~r^{\prime} > 5.4$ range, 
most of the post-merger galaxies seemed to be bulge-dominated regardless 
of their stellar masses. One massive galaxy showing a dusty disk and disturbed 
faint features may have been an obscured post-merger galaxy with high internal extinction
((b) in Figure 6).

We calculated the bulge-to-total (B/T) ratios for all the red sequence 
galaxies in \citet{she12} to verify that our post-merger classification 
had not been affected by the spiral structure of the late-type galaxies. To derive
the B/T ratios, we measured the radial surface brightness profiles of the galaxies 
using the ellipse task in IRAF. A least-squares 
fitting of the profiles was performed with a composite model of a de Vaucouleurs' profile 
and an exponential profile using IDL routines from the MPFIT package.
The best fit model was chosen using the minimum $\chi^{2}$ method. 
From the models, we calculated the B/T ratios of the galaxies. 
For more details on the B/T calculations, please refer to \citet{she12}. 

We compared the average B/T ratios of the red sequence galaxies as a function of 
the NUV $-~r^{\prime}$ colors and the M$_{r^{\prime}}$. Only the RSVL 
samples with $\widetilde{\chi}^{2} \le 2$ were included.
Before considering the B/T ratios, Figure~\ref{uvfrac} (a) shows the post-merger fractions 
among the RSVL galaxies in a NUV $-~r^{\prime}$ vs M$_{r^{\prime}}$ grid. 
It suggests that 1) the post-merger fractions of more massive galaxies are larger, 
and 2) all massive RSF galaxies 
(M$_{r^{\prime}} < -22$ and NUV $-~r^{\prime} \le 5$) show post-merger 
features. Figure~\ref{uvfrac} (b) shows the average B/T ratios for the 
post-merger galaxies. The post-merger samples were mostly
``bulge-dominated'' (B/T $>$ 0.4 for E/S0 galaxies following \citet{som99}). 
However, it should be noted that no robust profile fitting was obtained for 
about half of the UV-bright post-merger galaxies due to their 
asymmetric structures, as shown in Figure~\ref{cmd_uvthumb}.
The average B/T ratios for the normal galaxies are presented in Figure~\ref{uvfrac} 
(c). In general, their average B/T ratios were smaller than those of the post-merger
galaxies in the same bin. As shown in Table~\ref{avgfrac}, we found that the RSF 
fractions for normal, featureless red sequence galaxies were also large (up to 50\% with NUV $-~r^{\prime} \leq 5.4$). 
We considered the possibility that this may have been caused by passive spirals being included in optical red sequences. 
However, passive spirals did not seem to be the primary origin of those UV-bright  
normal galaxies, as their average B/T ratios were too large for them to be considered
spiral galaxies. Interestingly, ``disk-dominated" galaxies were found 
in the very red UV-optical color bins ($6 \leq$ NUV $-~r^{\prime} < 7$). These
most likely were late-type galaxies with high internal extinction.

\subsection{UV Upturn}

\begin{figure*}
\center{
\includegraphics[scale=0.55]{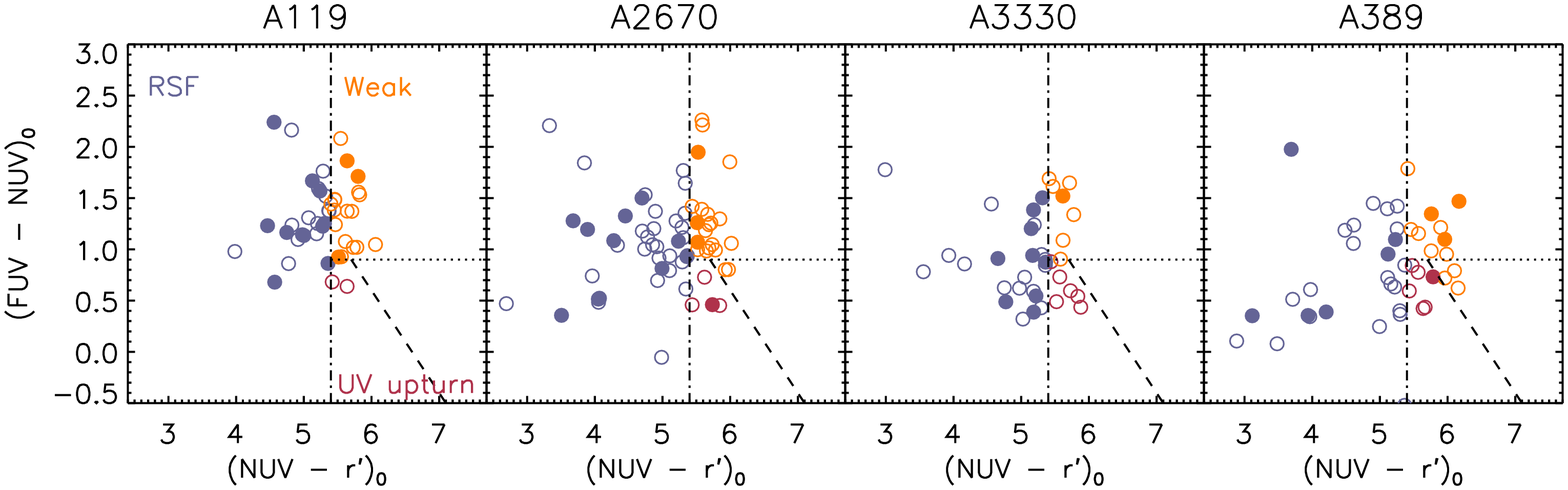}
\caption{UV-optical color-color diagrams of clusters. The demarcation lines 
used to divide the RSF, UV upturn, and UV weak regions were adopted from \citet{yi11}. 
The RSF galaxies are shown with blue symbols, while the UV upturn galaxies and UV weak 
galaxies are represented by red and orange symbols, respectively. Filled circles 
indicate post-merger galaxies. 
\label{uvccd} }
}
\end{figure*}

The UV-optical color-color diagrams were also examined to check the UV upturn phenomenon 
in the red post-merger galaxies. Among the RSVL samples, 189 galaxies 
(142 normal, 47 post-merger) were detected in both the FUV and NUV GALEX bands. 
The UV-optical color-color diagrams are presented in Figure~\ref{uvccd}. 
Demarcation lines were adopted from \citet{yi11} to select candidates for the UV upturn 
galaxies. About 4\% (2/47) of the post-merger galaxies were classified as UV upturn 
galaxies, while about 11\% (16/142) of the normal galaxies showed UV upturn features 
in this diagram (without correction from non-detections in the FUV). The smaller
UV upturn fraction among the post-merger galaxies may have been a consequence of the larger
RSF fractions among them, which hid the FUV excess from the evolved stellar systems. 

\citet{yi11} showed that UV upturn is not a common phenomena 
at $z <$ 0.1 using SDSS galaxy clusters. \citet{her14} suggested that binary 
star populations of very low metallicity can show UV-optical colors in the RSF regime. 
However a typical metallicity distribution would not expect a substantial fraction of metal-poor stars 
(e.g., \citet{kod97}). Therefore UV upturn is negligible in this study.

\subsection{Spatial Distributions}
\begin{figure}
\includegraphics[scale=0.5]{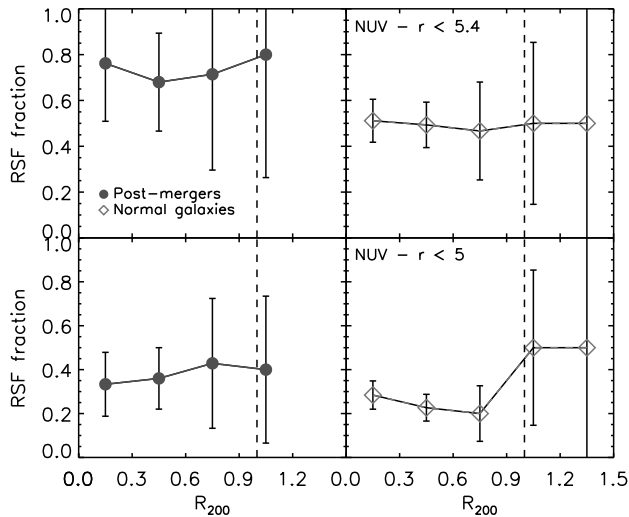}
\caption{RSF fractions against the clustocentric distance in units of R$_{200}$ for 
post-merger galaxies (filled circles) and normal galaxies (open diamonds). 
The bin size was 0.3 R$_{200}$. The upper panels show the RSF fractions for the
NUV $-~r^{\prime} = 5.4$ cut while the bottom panels present the results 
for NUV $-~r^{\prime} =5$. The dashed vertical line indicates 
$1\times$R$_{200}$.  \label{frac_dist}}
\end{figure}

Figure~\ref{frac_dist} presents the RSF fractions as functions of the clustocentric 
distance in R$_{200}$ units using RSF criteria of NUV $-~r^{\prime} = 5.4$ (upper panels)
and NUV $-~r^{\prime} = 5$ (bottom panels). The RSF fractions for the post-merger 
galaxies did not show any particular trend within R$_{200}$ under either of those RSF criteria.
This may indicate that post-merger RSF galaxies are randomly located within R$_{200}$ of 
a cluster, matching the results for the dependence of post-merger fractions on the
distance from the cluster center in \citet{she12}. In that study, it was shown that the fraction of 
UV-bright (NUV $-~r^{\prime} \leq 5$) normal galaxies was large at the cluster 
outskirts (R $>$ R$_{200}$). However, the number of galaxies at that distance in 
this study's sample was too small to be conclusive.

We performed a Dressler-Shectman test \citep{dre88} to identify sub-structures within 
the clusters. Sub-structures found in clusters may indicate that they have infalled
to a central cluster at relatively recent epoch. As a group environment induces 
more active evolutionary processes (e.g., galaxy mergers and star formation), in general, 
the correlation of the distribution of RSF galaxies with sub-structures is worthy of confirmation.
In short, the DS test was used to compare the local velocity and velocity dispersion 
for each galaxy with the global values. Using the spectroscopic members of each cluster, 
we first computed the mean velocity ($\bar{v}_{\rm cl}$) and the velocity dispersion ($\sigma_{\rm cl}$). 
The mean velocity $\bar{v}^{i}_{\rm local}$ and velocity dispersion $\sigma^{i}_{\rm local}$ of the 
10 nearest neighbors were also computed for each galaxy \textit{i}. These quantities were then combined 
to compute the individual galaxy deviations ($\delta_{i}$), as follows: 
\begin{equation} 
\label{delta_i}
\delta^{2}_{i}=\left(\frac{10+1}{\sigma^{2}_{\rm cl}}\right) \left[ (\bar{v}^{i}_{\rm local}  -  \bar{v}_{\rm cl})^{2} + (\sigma^{i}_{\rm local}  - \sigma_{\rm cl})^{2}  \right]
\end{equation}

\begin{figure*}
\center{
\includegraphics[scale=0.55]{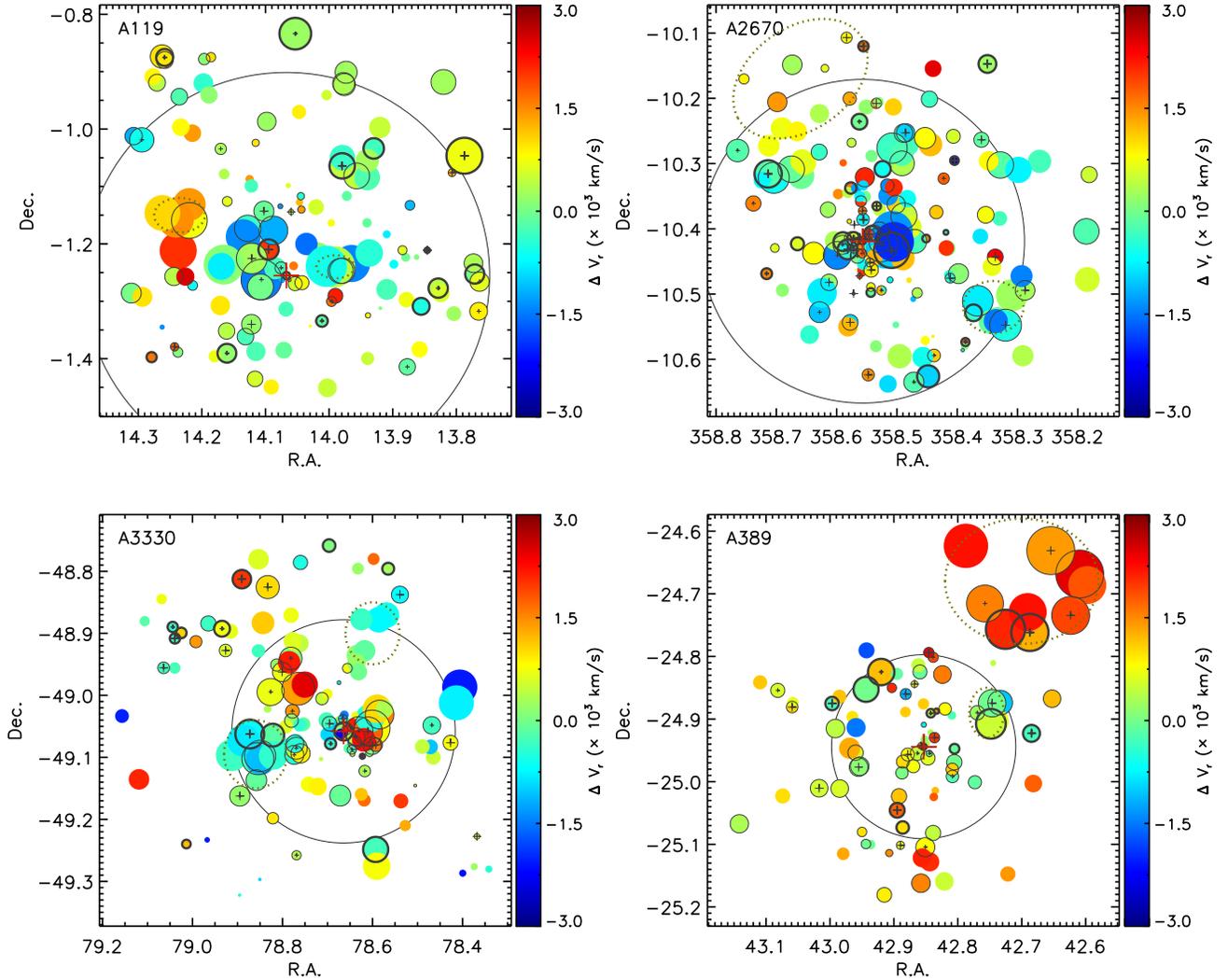}
\caption{Spatial distribution of all spectroscopic members in 
each cluster. The sizes of the filled circles show the $\delta i $ values from 
the Dressler-Shectman test and are color-coded according to their radial 
velocity centered at the median velocity of the cluster. The
possible sub-structures were carefully selected using galaxies with large $\Delta i$ values ($\Delta i \geq 2$), 
while also considering their spatial distributions and radial velocity histograms (dotted ellipses). 
The RSVL samples and RSF galaxies are represented with open gray circles and gray crosses 
in the center of the circles. The post-merger galaxies are represented with thicker gray symbols. 
The small crosses indicate galaxies with weak RSF signatures ($5 <$ NUV $-~r^{\prime} < 5.4$). 
The location of the BCG is marked with a red cross and a half-virial raius ($0.5\times$ R$_{200}$)
of each cluster is drawn with a circle in background. These plots suggest that the post-merger galaxies and 
RSF galaxies were not particularly related to the sub-structures. \label{dstest}}
}
\end{figure*}

\noindent The sub-structure candidates were initially identified using 
$\delta i$ values and were then confirmed through visual inspection of their spatial
and velocity distributions. If more than 4 galaxies with $\delta i > 2$ lay close to one another 
in terms of position and radial velocity, we selected them as a possible sub-structure.  
For more details, we refer readers to \citet{jaf13}. 
The distribution of the cluster members and possible group candidates is
presented in Figure~\ref{dstest}. 
Since we were looking mostly within R$_{200}$, we could not identify many robust sub-structures 
except for Abell 389, the furthest cluster sample, at $z = 0.112$. 
We inspected whether the RSF galaxies had any preference for lying in 
sub-structures.  As shown in Figure~\ref{dstest}, however,
neither post-merger galaxies nor post-merger RSF galaxies were preferentially found 
in the sub-structures.

\section{Summary and Discussion}
\label{sec:disc}

This study explored the GALEX UV properties of optical red sequence galaxies 
in 4 rich Abell clusters at $z \leq 0.1$. In particular, we tried to find a hint
of merger-induced recent star formation in the red sequence galaxies. 
Using the NUV $-~r^{\prime}$ colors of the galaxies, RSF fractions were
derived based on various criteria for post-merger galaxies and normal galaxies. 
Following $k-$correction, about 36\% of the post-merger galaxies were classified as 
RSF galaxies with a conservative recent star formation (RSF) criterion (NUV $-~r^{\prime} \leq 5$),
and that number was doubled ($\sim$ 72\%) when using a generous criterion 
(NUV $-~r^{\prime} \leq 5.4$). The trend was the same when the sample was limited
to galaxies within $0.5\times$R$_{200}$. 
Post-merger galaxies with strong UV emission were found to show more violent,
asymmetric features in the deep optical images. 
The RSF fractions did not show any trend along the clustocentric distance within R$_{200}$.
A Dressler-Shectman test was carried out to check whether the RSF 
galaxies had any correlation with sub-structures in the galaxy clusters. 
We could not identify many sub-structures from our clusters as the 
data only effectively included areas within R$_{200}$. Within our given field-of-view, 
the RSF galaxies did not appear to be preferentially related to the clusters' sub-structures.

The 30\% of RSF fraction among the volume-limited red sequence (RSVL) galaxies in 
galaxy clusters is what we have found through this study. The result is comparable with what \citet{kav07} 
found, at least 30\% of a RSF fraction among $\sim$2100 early-type galaxies across a wide range of environments. 
\citet{fit15} also looked into $\sim$29000 nearby early-type galaxies from SDSS and found that their star formation 
histories are mostly determined by structural parameters, not by environments.
These results suggest that currrent environment where an early-type galaxy sits in is not a critical factor
to determine its star formation history.

Although the RSF fraction of post-merger galaxies seemed larger
than that of normal galaxies, only 
30\% (21/70) of RSF galaxies among the optical red sequence galaxies showed
post-merger signatures in the deep optical images.
The remaining 70\% of UV-bright red sequence galaxies did not show 
morphologically-disturbed features, and they mostly populated the less massive end of the 
sample's mass range ($-21 \leq~$M$_{r^{\prime}} < -20$).  
It is possible that the study may have missed minor merger features in these galaxies 
from the deep optical images, as post-merger features become more difficult to detect
as the galaxy magnitude gets fainter. 
It has been claimed through a series of papers \citep{kav09,kav11,cro11,kav14} 
that minor merger has a significant role in the recent star formation in early-type galaxies.
Also Ji et al. (2014) reported that they could not see merger features from a 1:10 merger simulation 
with a given imaging depth which is comparable to ours.
However, considering the timescale difference for the fading of 
morphological merger features ($\sim$ 4 Gyr) and the young stellar populations 
($\sim$ 1 Gyr), the fraction of RSF red sequence galaxies without disturbed 
features was still significantly large. 
We also confirmed that those featureless RSF galaxies were mostly bulge-dominated early-type galaxies
and so they have a low chance to be a passively evolving spiral galaxy.
Our results imply that there may be other and perhaps more influential channels triggering
residual star formation in early-type galaxies. 

If recent galaxy merger is not a predominant driver of residual star formation 
in red sequence galaxies, another possible explanation 
may be gas cooling from internal processes.
The source of cold gas for early-type galaxies remains an open question.
According to theoretical studies \citep{kim11,lag14,vij15}, external sources may be gas accretion from minor mergers and
internal processes may be stellar mass-loss or hot gas cooling. 
\citet{lag14} concluded by comparison of semi-analytic models to observations of 
gas in early-type galaxies that more than 90\% of neutral gas contents were supplied by radiative cooling 
from hot haloes, 8\% by gas accretion from minor mergers, and 2\% by mass-loss from old stars.
\citet{vij15} simulated ram pressure stripping of hot gas from galaxies in cluster environment
and suggested that stripped gas can be confined in a galaxy's gravitational potential in the form of tail.
Therefore, it is possible that the 70\% of featureless RSF galaxies in this study may be mainly fuelled by their hot
haloes.  

According to the result, it seems that post-merger feature in low surface brightness 
is not a critical hint for RSF in red sequence galaxies. Therefore still NUV excess remains as a powerful 
tool to find RSF in early-type galaxies effectively.

This study is focused on the photometric properties of galaxies. 
A future study that makes use of spectroscopic data with adequate 
signal-to-noise ratios will be useful to address the effect of weak AGN
on RSF in elliptical galaxies, as suggested by \citet{val15},
and to derive various star-formation related properties.

\acknowledgments

We thank the anonymous referee for his/her constructive comments and suggestions. 
SKY acted as the project leader and a corresponding author.
YKS acknowledges support from FONDECYT Postdoctoral Fellowship 
(No. 3130470) by CONICYT. SKY acknowledges support from the 
Korean National Research Foundation (NRF-2014R1A2A1A01003730). 
YJ acknowledges support from the Marie Curie Actions of the European 
Commission (FP7-COFUND). 
RD gratefully acknowledges the support provided by the BASAL Center 
for Astrophysics and Associated Technologies (CATA), 
and by FONDECYT grant No. 1130528. ET acknowledges support from 
CONICYT Anillo ACT1101.



{\it Facilities:} \facility{GALEX, Blanco (MOSAIC 2 CCD Imager)}

\appendix
\section{A Table of RSF fractions}
\label{append}

\def\arraystretch{1.5}
\begin{deluxetable*}{llllll}
\tabletypesize{\footnotesize}
\tablewidth{0pt}
\tablecaption{RSF Fractions2 (in percent)\tablenotemark{a} \label{fraction}}
\tablehead{
\colhead{Samples} & \colhead{Cluster ID} & \multicolumn{2}{c}{w/ k-correction} & \multicolumn{2}{c}{w/o k-correction} \\ 
\hline
\colhead{} & \colhead{} & \colhead{NUV $-~r^{\prime} \leq$ 5.4} & \colhead{NUV $-~r^{\prime} \leq$ 5} & \colhead{NUV $-~r^{\prime} \leq$ 5.4} & \colhead{NUV $-~r^{\prime} \leq$ 5}
}
\startdata
Post-merger&A119&76.9 (10/13) &38.5 (5/13)&69.2 (9/13) &30.8 (4/13)  \\
&A2670&72.2 (13/18) &50.0 (9/18)&66.7 (12/18) &44.4 (8/18)  \\
&A3330&78.6 (11/14) &21.4 (3/14)&57.1 (8/14) &21.4 (3/14)  \\
&A389&61.5 (8/13) &30.8 (4/13) &53.8 (7/13) &30.8 (4/13) \\
\hline
&Average&72.4 (42/ 58) &36.2 (21/ 58)&62.1 (36/ 58) &32.8 (19/ 58)  \\
\hline
Normal galaxies&A119&40.0 (16/40) &17.5 (7/40)&32.5 (13/40) &15.0 (6/40)  \\
&A2670 &54.0 (34/63) &33.3 (21/63)&44.4 (28/63) &25.4 (16/63) \\
&A3330&53.8 (21/39) &25.6 (10/39) &48.7 (19/39) &25.6 (10/39) \\
&A389&50.0 (22/44) &25.0 (11/44)&40.9 (18/44) &22.7 (10/44)  \\
\hline
&Average&50.0 (93/186) &26.3 (49/186)&41.9 (78/186) &22.6 (42/186)  \\
\hline
Post-merger&A119&72.7 (8/11) &36.4 (4/11) &63.6 (7/11) &36.4 (4/11) \\
($<$ 0.5 R$_{200}$)&A2670&68.8 (11/16) &43.8 (7/16)&62.5 (10/16) &37.5 (6/16)  \\
&A3330&100 (4/4) & 0.0 (0/4) &75.0 (3/4) & 0.0 (0/4) \\
&A389&55.6 (5/9) &11.1 (1/9) &44.4 (4/9) &11.1 (1/9) \\
\hline
&Average&70.0 (28/ 40) &30.0 (12/ 40) &60.0 (24/ 40) &27.5 (11/ 40) \\
\hline
Normal galaxies&A119&42.9 (15/35) &20.0 (7/35) &37.1 (13/35) &17.1 (6/35) \\
($<$ 0.5 R$_{200}$)&A2670&53.8 (28/52) &32.7 (17/52)&46.2 (24/52) &25.0 (13/52)  \\
&A3330&56.5 (13/23) &21.7 (5/23)&52.2 (12/23) &21.7 (5/23)  \\
&A389 &51.7 (15/29) &24.1 (7/29)&37.9 (11/29) &20.7 (6/29) \\
\hline
&Average &51.1 (71/139) &25.9 (36/139) &43.2 (60/139) &21.6 (30/139)\\
\hline
All RSVL&Average&55.3 (135/244) &28.7 (70/244)&46.7 (114/244) &25.0 (61/244)  \\
All RSVL ($<$ 0.5 R$_{200}$) &Average&55.3 (99/179) &26.8 (48/179)&46.9 (84/179) &22.9 (41/179) 
\enddata
\tablenotetext{a}{Galaxy counts are presented in the parentheses.}
\end{deluxetable*}

\end{document}